# EFFECT OF $Ta^{5+}$ SUBSTITUTION ON THE GROUND STATE OF CMR MANGANITES

L. Seetha Lakshmi[a,b,*], K. Dörr[b], K. Nenkov[b], A. Handstein[b], V. Sridharan[a], V. S. Sastry[a] and K. -H. Müller[b]

[a]Materials Science Division, Indira Gandhi Centre For Atomic Research, Kalpakkam, 603102, India
[b]IFW Dresden, Postfach 270116, Dresden 01171, Germany

(21.12.2005)

**Abstract**

For the first time, we report the dominant role of charge state in modifying the ground state of CMR manganites. $Ta^{5+}$ substitution results in an appreciable increase in lattice parameters and unit cell volume due to increased $Mn^{3+}$ concentration. The ferromagnetic-metallic ground state modifies to a cluster glass insulator for x > 0.03. The reduction in the transition temperatures with increasing x is ~39K / at.%, the largest reported for the Mn site substituted CMR manganites. Two distinct magnetic transitions, thermomagnetic irreversibility, non-saturation in magnetization, non-stationary dynamics and a broad maximum in magnetic specific heat confirm *a cluster glass state* for x = 0.10.

## 1. Introduction

In manganites, the strong interplay between charge, spin, lattice and orbital degrees of freedom leads to a variety of phases with distinct ground state properties. As the essential degrees of freedom are intimately linked to Mn ion, Mn site substitution offer a direct probe to elucidate its physical properties Earlier works[1-3] show that all the substituents in ferromagnetic metallic manganites, irrespective of their chemical nature lower the ferromagnetic transition temperature, but to different extents. In our previous studies on Mn site substitutions with both diamagnetic and paramagnetic ions, we rationalized the variation of extent of suppression in the transition temperatures with concentration ($dT_c/dx$) in terms of local structural modification due to size mismatch and local magnetic coupling between the magnetic moments of substituents and Mn ion[4]. In this paper, we address the role of yet another important parameter, the valence state of the substituent in modifying its ground state properties. With this notion, for the first time, we report the effect of pentavalent Ta substitution at the Mn site of $La_{0.67}Ca_{0.33}MnO_3$: Diamagnetic ($4d^{10}$) $Ta^{5+}$ ion having its ionic radius (0.640 Å) very close to that of $Mn^{3+}$ ion (0.645 Å)[5] is expected to introduce negligible structural modification due to size effect. Also by virtue of its closed shell configuration, $Ta^{5+}$ ion does not introduce any additional magnetic coupling effects. From our present studies, we show that charge state (valence) of the substitution has a dominant role in modifying the ground state of colossal magnetoresistive manganites.
.

* Corresponding author. Tel.: +91-4114-27480347
E-mail address: slaxmi73@gmail.com

## 2. Experiment:

Polycrystalline compounds were synthesized by solid-state reaction. The final sintering of the samples at 1500ºC was carried out in a single batch to ensure identical sintering conditions. The high statistics room temperature powder XRD patterns were recorded in the 2θ range, 15-120° using CuKα radiation (STOE). The detailed analysis of the crystal structure was carried out using GSAS[6] Rietveld refinement program and SPUDS[7] program to generate the starting model for the Rietveld analysis. The temperature variation of resistivity was measured by conventional four probe method. A split coil superconducting magnet was employed for steady magnetic fields up to 7 T with magnetic field parallel to the current. The temperature variation of ac susceptibility ($\chi'(T)$) under a field of 1Oe and a frequency of 133Hz was measured using a Lakeshore 7000 series susceptometer. The frequency and field dependent $\chi'(T)$ was carried out in the zero field cooled (ZFC) condition using a physical property measurement system (PPMS, Quantum Design). ZFC and field cooled (FC) thermomagnetization, field dependence of magnetization and the time dependent decay of the remanent magnetization were recorded using a SQUID magnetometer (Quantum Design). The specific heat measurements in the temperature range 300K- 2K were performed under ZFC condition in the PPMS and data were collected during the warming cycle.

## 3. Results and Discussion

$La_{0.67}Ca_{0.33}Mn_{1-x}Ta_xO_3$ (0 ≤ x ≤ 0.10) compounds are single phase with orthorhombic structure and

P*nma* symmetry[8]. The lattice parameters and average Mn-O distance ($d_{Mn-O}$) show a systematic increase with Ta substitution (Fig1a and Fig.1c, hence an overall expansion of the unit cell is observed. (Fig1b). On the other hand, average Mn-O-Mn bond angle (<Mn-O-Mn>) (Fig.1d) shows a considerable decrease in the entire range of substitution. Furthermore, a substantial drop of $Mn^{4+}$ concentration is envisaged with an increase in x [Fig.1f].

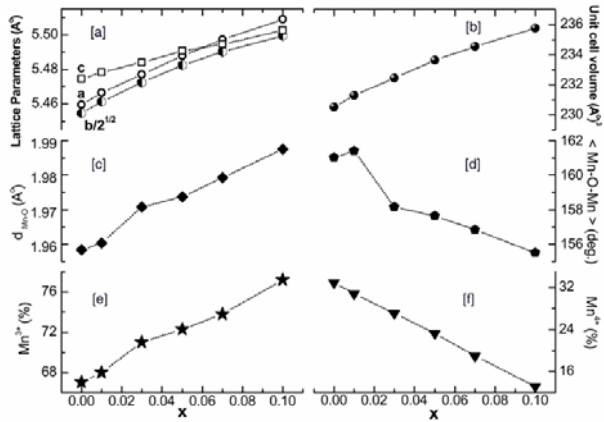

**Figure :1** Variation of (a) lattice parameters (a, b and c) (in Å) (b) unit cell volume ( in Å$^3$) (c) $d_{Mn-O}$ (in Å) (d) <Mn-O-Mn> ( in deg.) (e) $Mn^{3+}$ concentration (%) (f) $Mn^{4+}$ concentration(%) as a function of x of $La_{0.67}Ca_{0.33}Mn_{1-x}Ta_xO_3$ ($0 \leq x \leq 0.10$) compounds.

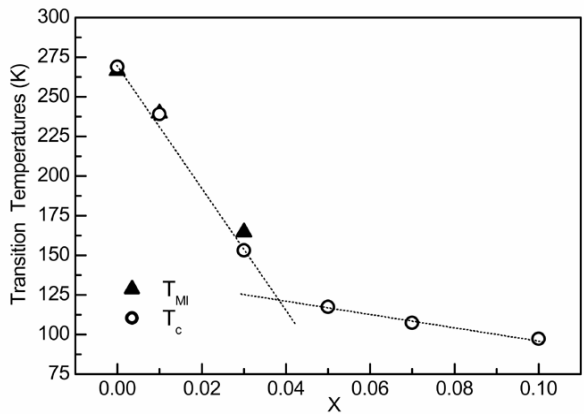

**Figure :2** Variation of metal-insulator transition temperature ($T_{MI}$) and Curie temperature ($T_c$) as function of x of $La_{0.67}Ca_{0.33}Mn_{1-x}Ta_xO_3$ ($0 \leq x \leq 0.10$) compounds. Solid line is the best linear fit to the experimental data to estimate the rate of suppression in the transition temperatures.

The compounds with $x \leq 0.03$ exhibit a metal to insulator transition at $T_{MI}$ and close to it, a para to ferromagnetic transition at $T_c$ is also observed[8]. Beyond x = 0.03, the compounds exhibit an insulating behavior with no perceptible anomaly near $T_c$, indicating the non-metallic behavior of the ferromagnetic phase. It is worth mentioning that *the reduction in $T_c$ as well as $T_{MI}$ is about ~39K/at.%, the largest extent reported for the Mn site substitutions in CMR manganites* (Fig.2). While $T_{MI}$ decreases at a constant rate of ~36K/at.%, a linear decrease of similar rate (~39K/at.%) is observed for $T_c$ up to x=0.03 , which levels off to a much smaller rate (10K/at.%) for higher Ta concentration. Under an applied magnetic field of 7 T, the compounds with x $\leq$ 0.03 exhibit a significant reduction of ρ with a shift of $T_{MI}$ to higher temperatures, a distinct feature of CMR manganites. For x > 0.03 no field driven metallic state could be discerned at low temperatures even at the presence of 7 T. Another striking feature of x > 0.03 is the presence of a cusp like feature in $\chi'(T)$ below $T_c$, followed by a broader shoulder at lower temperatures[8]. These results indicate that the Ta substitution modifies the ground state to a *glassy insulator*. For further understanding of the modified magnetic ground state of these compounds, we have carried out a detailed investigation of x = 0.10, the highest substitution level of the present study.

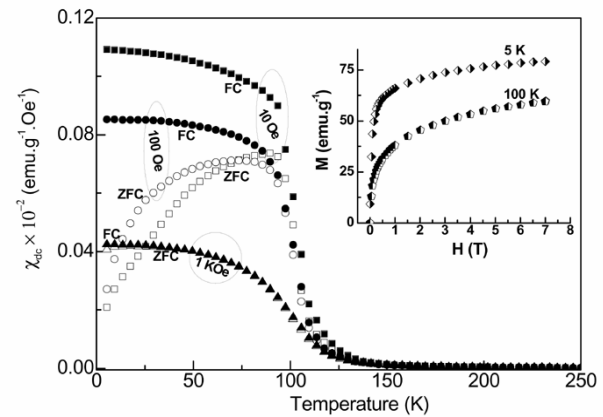

**Figure :3** ZFC and FC thermomagnetization curves of $La_{0.67}Ca_{0.33}Mn_{0.90}Ta_{0.10}O_3$ compound for H=10Oe, 100Oe and 1 KOe. Inset shows the field dependence of magnetization at 100K and 5 K.

Fig.3 shows the dc susceptibility ($\chi_{dc}(T)$) as a function of the applied magnetic field (H = 10 Oe,100 Oe and 1 KOe) for x = 0.10 in both ZFC and FC conditions. A broad cusp-like anomaly is found at a temperature just below $T_c$ in ZFC $\chi_{dc}(T)$ when the applied field is 10 Oe, but this cusp loses its sharpness and becomes a broad maximum with increasing magnetic field up to 1 KOe. An important point to be noted here is that the values of FC measurement of $\chi_{dc}(T)$ continue to increase strongly below the irreversibility temperature $T_r$ (just below $T_c$), at which the ZFC and FC curves merge, a typical feature of various cluster glass systems[9]. On the other hand in canonical spin glass systems, the FC $\chi_{dc}(T)$ shows a nearly constant value below $T_r$. It is noteworthy that due to the fact that cluster-glass exhibit short range intra-cluster ferromagnetism below $T_c$, it may mimic some features of those found in re-entrant spin glass system (RSG). Nevertheless, it may be noted that in several RSG systems, $T_r$<< $T_c$[10], whereas in a cluster glass, the irreversibility arises just below $T_c$, as observed in Fig. 3 for x=0.10. Additionally, strong reduction of the thermomagnetic irreversibility with enhancing measuring field indicates a field induced short range of inter-cluster ferromagnetic ordering below $T_r$. The absence of long range ordering is also apparent from the non saturation of magnetization even up to a field of 7 T even at 5 K (inset of Fig 3).

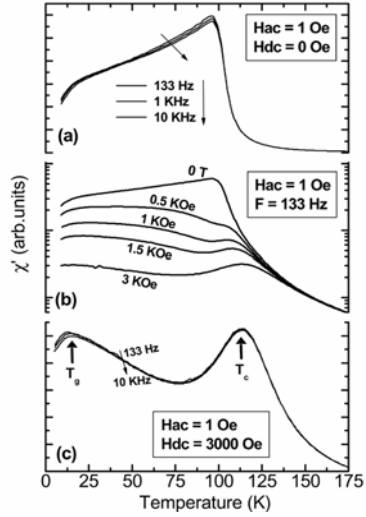

**Figure :4** Temperature variation of ac susceptibility of $La_{0.67}Ca_{0.33}Mn_{0.90}Ta_{0.10}O_3$ compound measured in
(a) different frequencies (f) and in a ac field (h)of 1 Oe
(b) different static bias fields with F=133 Hz and h=1Oe
(c) different F and h= 1Oe in 3KOe static magnetic field.

In order to probe the dynamics of spin freezing, $\chi'(T)$ was measured as a function of frequency and static bias field. In the absence of static bias field $\chi'(T)$ shows a frequency dependent cusp just below $T_c$ and a broad shoulder at lower temperatures (Fig.4a). Since the ferromagnetic transition smears out the cluster-glass transition, it is difficult to identify the spin freezing temperature. The dynamic magnetic response gets modified drastically with an application of superimposed dc magnetic field (Fig.4b). The suppression of $\chi'(T)$ and the broadening of the transition are due to the absence of long range ferromagnetic order as well as the slow response of the clusters in the presence of dc field. When the field is further increased, two distinct magnetic transitions, marking the ferromagnetic transition at $T_c$, followed by a spin freezing transition at $T_g(H)$ were found. With increasing magnetic fields, the decline of $\chi'(T,H)$ below 12 K and the shift of $T_g(H)$ to lower temperatures together confirm the presence of frozen spin disordered state. The shift in the susceptibility peak is not distinct in the absence of static bias field, whereas it is clearly seen at $T_g(H)$ with an increase in frequency (133Hz ≤ F ≤ 10 KHz) in the presence of 0.3 T field (Fig.4c). The frequency dependent shift of $T_g(H)$ to higher temperature is one of the distinct features of cluster glass. It is worth mentioning that there is no frequency dependent shift of the peak at $T_c$.

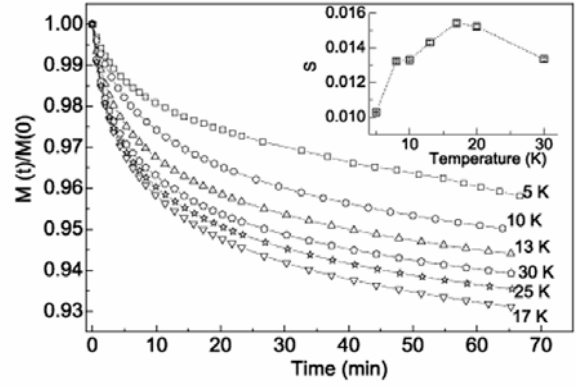

**Figure :5** Decay of remanent magnetization as a function of time of $La_{0.67}Ca_{0.33}Mn_{0.90}Ta_{0.10}O_3$ compound Dotted line is the best fit to the experimental data. Inset shows magnetic visocity (S) as a function of temperature refer text for details.

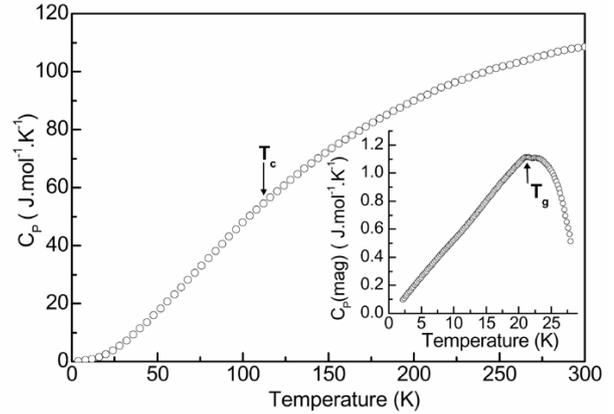

**Figure : 6** Temperature dependence of specific heat ($C_p$) of $La_{0.67}Ca_{0.33}Mn_{0.90}Ta_{0.10}O_3$ compound. Inset shows the magnetic contribution to the total specific heat after subtracting the lattice contribution. Refer text for details.

Non-equilibrium spin dynamics is another characteristic feature of the cluster glass. The decay of remanent magnetization (M(t)) was measured in the temperature range 5K - 90K( Fig.5) A best fit using logarithmic relaxation of the functional form $M(t)= M(0) (1-Sln(t/t_0))$ (where S denotes the magnetic viscosity) is observed not only below but also above $T_g(H)$. The clear maximum of S(T) close to $T_g(H)$ (inset of Fig.5) results from the competition between two processes (i) freezing of magnetic moments due to the presence of competitive magnetic interactions (ii) activation of the frozen-in moments with increasing temperature. To ascertain the cluster glass state at zero field, specific heat measurements ($C_p(T)$) were carried out in the temperature range 2-300K. No distinct anomaly corresponding to $T_c$ could be found in $C_p(T)$ (Fig.6). In order to estimate the magnetic specific heat, the low temperature specific heat (2-25K) is fitted by the following equation. $Cp(T) = \beta T^3 + \alpha T^5 + \delta T^n$ where $\beta T^3 + \alpha T^5$ represent the lattice contribution and $\delta T^n$ is the spin wave contribution ($C_m$) to the specific heat. The value of the exponent n depends on the nature of the excitations. Since the compound is highly insulating, there is no electronic contribution to the total specific heat. The Debye temperature ($\theta_D$) is estimated to be $447 \pm 5.89$ K and is comparable with the reported values for manganites in the literature[11]. The value of n is estimated to be $1.04 \pm 0.01$, which is close to the glassy linear specific heat. The magnetic specific heat ($C_m$) (inset of Fig.6) is obtained after subtracting the phonon contribution. Linear temperature dependence of $C_m$ together with a broad maximum just above $T_g$ provide a strong evidence for the cluster-glass state.

In the following, the microscopic consequences of $Ta^{5+}$ substitution are discussed. Taking charge neutrality into account, $Ta^{5+}$ is expected to strongly shift the average valence state of Mn towards 3+, according to $La^{3+}_{0.67}Ca^{2+}_{0.33}Mn^{3+}_{(0.67+x)}Mn^{4+}_{(0.33-2x)}Ta^{5+}_xO_3$. Thus, the appreciable increase of the lattice parameters with $Ta^{5+}$ substitution (Fig.1(a)) can be understood as a result of the larger average ionic radius at the Mn site, though electron-lattice coupling might give an additional influence. Further, there is a substantial drop of the carrier density ($Mn^{4+}$ concentration) with increasing x (Fig1(f)). Indeed, the consequences of these two effects (reduced carrier density and increased ionic radius at the Mn site) should be displayed in similar way by compounds $La_{1-z}Ca_zMnO_3$[12] if the same level of doping ($Mn^{4+}$ content) is compared. Then, Mn site ionic radius is also nearly equal for respective compounds, while the La site ionic radius remains nearly constant due to similar radius of La and Ca. However, the Ta substituted samples exhibit much lower values of $T_c$. For instance for x = 0.05, 23 % of Mn sites contain $Mn^{4+}$ ions (or 21.9 % of present Mn ions are $Mn^{4+}$), leading to $T_c = 117$ K and low temperature glassy behavior. On the other hand, ferromagnetism and metallicity are found in $La_{0.77}Ca_{0.23}MnO_3$ ($T_C \sim 220$ K) with same doping level[11]. Therefore, the effects of carrier density and ionic size seem insufficient to account for the observed strong suppression of ferromagnetism. Larger structural modification involving an increase in the average Mn-O bond length ($d_{Mn-O}$) and a substantial decrease in Mn-O-Mn bond angles ($<$Mn-O-Mn$>$) has been detected for Ta substituted compounds which is expected to decrease the DE interaction strength. The diamagnetic $Ta^{5+}$ ion ($4d^{10}$), by virtue of its closed shell configuration, just dilutes the magnetic network, and dilution generally results in a reduced magnetic ordering temperature. However, the degree of dilution is rather low in the studied compounds, and other diamagnetic substitutions for Mn affect $T_c$ much less than Ta. Hence, another microscopic effect seems necessary to understand the experimental result. As pointed out by Alonso $et. al.$[13], the changes in charge state at the Mn site may create a random local electrostatic potential that attracts or repels charge carriers (holes). Taking the local electrostatic potential into account, calculations of these authors (carried out for $Ga^{3+}$ ions) indicate enhanced suppression of $T_C$ and glassy states. Due to the large positive charge of $Ta^{5+}$ ion, one would expect a strong electrostatic effect for this substitution, though the potential is positive in the case of $Ta^{5+}$. We suggest that this mechanism might contribute to the observed unusually strong suppression of the itinerant DE ferromagnetism. In the presence of antiferromagnetic superexchange interactions and spatial disorder, the appearance of glassy behavior is not unusual in manganites. The ability of $Ta^{5+}$ ion to induce cluster glass and insulating behavior at very low Ta concentration (x ~ 0.05) if compared to other trivalent and tetravalent substitutions for Mn underlines the significance of charge state (valence) of the substituent in modifying the ferromagnetic-metallic ground state of CMR manganites.

**Acknowledgement:**

Support by DFG, FOR520 is gratefully acknowledged. LSL also thanks CSIR, India for a Senor Research Fellowship.

**Reference:**

[1] J. Blasco, J. Garcia, J. M. de Teresa, M. R. Ibarra, J.Perez, P. .A. Algarabel, C. Marquina, C. Ritter, Phys. Rev.B 55 (1997), 8905

[2] Young Sun, Xiaojun Xu, Lei Zheng, Yuheng Zhang, Phys. Rev. B 60 (1999) 12 317.

[3] Xianming Liu, Xiaojun Xu, Yuheng Zhang, Phys. Rev. B 62 (2000) 15 112

[4] L. Seetha Lakshmi, V. Sridharan, D.V. Natarajan, V. Sankara Sastry, T.S. Radhakrishnan, Pramana, J. Phys. 58 (2002)1019.\

[5] R.D. Shannon, C.T.Prewitt , Acta Crystallogr.Sec. A 32 (1976) 751

[6] A.C.Larson, R.B. Von Dreele; Los Alamos National Laboratory Report LAUR 86-748 (2000)

[7] M.W.Lufaso , P.M Woodward, Acta Cryst., B 57 ( 2001) 725.

[8] L.Seetha Lakshmi, K.Dörr, K.Nenkov, V.Sridharan, V.S.Sastry and K.-H.Müller,cond-mat/0410327and J. Magn. Magn. Mater, (2005) ( To appear)

[9] D.A. Pejakovic, J. L. Manson, J.S.Miller and A.J. Epstein, Phys. Rev. Lett. 85 (2000) 1994

[10] S. N.Kaul, S. Srinath J. Phys. Condens. Matter 10 (1998) 11067

[11] L. Ghivelder, I. A. Castillo, N. McN.Alford, G. J. Tomka, P. C. Riedi, J.M. Driscoll, A. K. M. Hossain, L .F. Cohen, J.Magn.Magn.Mater, 189 (1998) 274

[12] P.Schiffer, A.P.Rameriz, W.Bao, S.-W.Cheong, Phys.Rev.Lett 75 (1995) 3336

[13] J. L. Alonso, L.A. Fernandez , F.Guinea, V. Laliena, V. Martin-Mayor, Phys. Rev. B 66, 104430 (2002)